\documentstyle[preprint,aps]{revtex}
\begin{document}
\draft
\title{NONINTEGRABLE INTERACTION OF ION ACOUSTIC AND ELECTROMAGNETIC 
WAVES IN A PLASMA}
\author{F.B. Rizzato$^1$, S.R. Lopes$^2$, and A.C.-L. Chian$^3$}
\address{$^1$Instituto de F\'{\i}sica - Universidade Federal do Rio Grande 
do Sul, P.O. Box 15051, 91501-970 Porto Alegre, Rio Grande do Sul, Brazil}
\address{$^2$Departamento de F\'{\i}sica - Universidade Federal do Paran\'a, 
P.O. Box 19081, 81531-990 Curitiba, Paran\'a, Brazil}
\address{$^3$National Institute for Space Research-INPE, P.O.Box 515, 12227-010 S\~ao 
Jos\'e dos Campos, S\~ao Paulo, Brazil}
\maketitle
\date{\today}
\begin{abstract}
{\it 
\noindent In this paper we re-examine the one-dimensional interaction of 
electromagnetic and ion acoustic waves in a plasma. Our model is similar to one solved 
by Rao {\it et al.} (Phys. Fluids, {\bf 26}, 2488 (1983)) under a number of 
analytical approximations. Here we perform a numerical investigation to examine the 
stability of the model. We find that for slightly over dense plasmas, the propagation 
of stable solitary modes can occur in an adiabatic regime where the ion acoustic 
electric field potential is enslaved to the electromagnetic field of a laser. 
But if the laser intensity or plasma density increases or the laser 
frequency decreases, the adiabatic regime loses 
stability via a transition to chaos. New asymptotic states are attained when the 
adiabatic regime no longer exists. In these new states, the plasma becomes rarefied, 
and the laser field tends to behave like a vacuum field. 
 }
\end{abstract}

-------------------\\
{\scriptsize PACS: 52.35.Mw; 05.45.+b}
 
\newpage
\section{Introduction}

The recent developments of new laser technologies allow the creation of strong 
pulses that can propagate in a plasma, either to accelerate particles \cite{nob85} or 
induce laser assisted fusion \cite{kru88}. Now, if a laser pulse interacts with a 
plasma the possibility exists of nonlinear wave coupling involving the pulse itself 
and nonlinear plasma modes \cite{kru88,wil77}. Since the variety of these nonlinear 
plasma modes is large, and since each mode exhibits a considerable richness regarding 
amplitude, polarization and frequency range \cite{chi96,lop96}, the nonlinear mode 
coupling is a feature to be appreciated with care. 

One important example of nonlinear mode coupling concerns the interaction of 
relativistically strong electromagnetic waves with Langmuir waves 
\cite{chi75,chi81,kaw83}. These analysis are restricted to simple wave  
solutions with well defined propagation velocities which are superluminous in an 
unmagnetized plasma. For these superluminous waves it is shown that the wave dynamics 
can be viewed as similar to the coupled dynamics of two nonlinear oscillators with 
their natural frequencies given by real numbers. One of the oscillators describes the 
transverse electromagnetic field and the other the longitudinal electrostatic field. 
The dynamics is found to be predominantly integrable. Some nonlinear resonant island 
chains are present in the appropriate Poincar\'e maps \cite{kaw83}, but their overlap 
is so small that the resulting trajectories are mostly regular.  

In another important range of subluminous wave velocities, around the ion acoustic 
range, the laser wave can couple to ion acoustic plasma modes. In this case the 
coupled waves propagate with velocities close to the ion acoustic velocity which is 
much smaller than the velocity of light. Here the ionic dynamics plays a crucial role 
and the resulting dynamics, in principle, bears no resemblance to the dynamics 
described in the previous paragraph. A good deal of analytical work has been done in 
Ref.\cite{rao83} to show that if one assumes once more a common and constant 
propagation velocity, the system becomes again equivalent to a pair of nonlinearly 
coupled oscillators. A novel feature studied here is that when the laser frequency 
is smaller than the average plasma frequency, wave localization can occur such that 
one has the formation of intense solitary pulses. In terms of the system of 
equivalent oscillators, the possibility of solitary pulses occurs when by varying the 
appropriate parameters the linear frequency of that oscillator describing the laser 
wave becomes imaginary. Using the language of nonlinear dynamics, this change in the 
character of the frequency occurs because the central elliptic point present in the 
appropriate phase space turns into a hyperbolic point \cite{lili83}.  

The work by Rao {\it et al.} \cite{rao83} utilizes powerful approximation techniques, 
but is essentially analytical as commented before. Therefore one would like to have 
some information on the stability of the solitary pulses thus formed, and this 
is what we do here. The stability issue has been already addressed several times over 
the past years, but focus has been preferentially directed upon the linear perspective 
\cite{inf90}. However, if one wishes to have some additional insight on the nonlinear 
development of these instabilities, the most appropriate tools of investigation appear 
to be the Poincar\'e maps mentioned above. With these maps one records the phase space 
coordinates of one of the oscillators, as one of the coordinates of the other crosses 
its zero with a definite sign for the derivative. Since the system is two degrees of 
freedom and since there exists a conserved Hamiltonian for the system, the point 
recorded on the map gives all the relevant information for the dynamics. Regular 
motion is associated with smooth curves of the Poincar\'e maps and chaotic (or 
nonintegrable) motion is associated with an erratic distribution of points 
representing the trajectory. We find here that the stability of the solitary pulses is 
quite limited. In fact we found no regular motion for velocities sufficiently below 
the ion acoustic velocities and for effective laser frequencies sufficiently below the 
average plasma frequency. In those chaotic situations the final asymptotic behavior 
looks like the one corresponding to uncoupled waves in the vacuum. The intrinsic 
instability of the system for too low values of the effective laser wave frequency, in 
particular, seems to preclude the formation of some special soliton solutions as we 
shall see. Chaos is present because our system is two degrees of freedom, and the 
system is two degrees of freedom because we allow for charge separation in the ion 
acoustic dynamics. Had we used quasineutrality assumptions, the system would be one 
degree of freedom and chaotic solutions would be absent. We finally point out that the 
transition to chaos we detect here is of a conservative character, so the chaotic 
dynamics we expect to see is of different type from the one present in dissipative 
systems \cite{lop96,ott80}.

The paper is organized as follows: in \S II we review the basic theory governing the 
interaction of a laser and an ion acoustic mode; in \S III we analyze the nonintegrable 
dynamics with help of Poincar\'e maps and in \S IV we conclude the work. 

\section{The Basic Theory}

\subsection{Introductory Remarks}

We consider here the interaction of a laser wave and an ion acoustic mode in a 
globally neutral plasma consisting of light electrons and massive ions. Let us 
re-derive the governing equation for the laser mode in a slightly different way from 
the one used in Ref. \cite{rao83}. If one assumes that the waves are plane waves 
propagating along the $z$ axis, the equation governing the high-frequency laser 
dynamics can be written in the form 
\begin{equation}
{\partial^2 \Psi \over \partial \, t^2} - c^2 {\partial^2 \Psi \over \partial \, z^2} = 
-{4 \pi q^2 \over m_e} (n_o + \delta n_e) {\Psi \over \sqrt{1+|\Psi|^2}},
\label{eqpsidim}
\end{equation}
where $-q$ is the electron charge, $m_e$ its rest mass, $n_o$ the 
average density, $\delta n_e$ the fluctuations of the electronic density due 
to the action of the waves, and $c$ the velocity of light. In Eq. (\ref{eqpsidim}) the 
laser intensity is considered strong enough to drive electrons to relativistic 
velocities. The field $\Psi (z,t)$ is defined in terms of the relation 
\begin{equation}
{q {\bf A} \over m_e c^2} = {1 \over 2} (\hat {\bf x} -i \hat {\bf y}) \, \Psi 
+ c. c.,
\label{defpsi}
\end{equation}
where ${\bf A}$ is the vector potential of a circularly polarized laser field, 
with $i^2 = -1$ and $c.c.$ designating complex conjugate. 
Now we assume solutions of the form 
\begin{equation}
\Psi(z,t) = \psi(\xi) \, e^{i(k z - \omega t)} 
\label{psi}
\end{equation}
with $k$ and $\omega$ respectively as the wave vector and effective frequency of the 
laser, with $\xi \equiv z-Vt$, and $\psi$ as a real slowly varying variable; $V$ 
denotes the propagation velocity. Next we substitute relation (\ref{psi}) into 
Eq. (\ref{eqpsidim}) and separates the resulting equation into its real and imaginary 
parts. The imaginary component yields the relation  
\begin{equation}
V = {c^2 k \over \omega},
\label{vwk}
\end{equation}
from which one can determine the propagation velocity, given the frequency and 
wave vector of the carrier. The real part, on the other hand, yields a governing 
equation for the real amplitude $\psi$, valid when $V \ll c$;
\begin{equation}
\beta \, {d^2 \psi \over d \xi^2} = - \Delta \, \psi + 
(1 + \delta n_e) \, {\psi \over \sqrt{1+\psi^2}}.
\label{psinor}
\end{equation}
In the equation above, space is normalized by the electronic Debye length 
$\lambda_{De}=(T_e/4 \pi n_o q^2)^{1/2}$, time by the ion plasma frequency 
$\omega_{pi}^2 = 4 \pi n_o q^2 / m_i$ and the density by $n_o$; $m_i$ is the ion 
mass, $T_e$ is the electron temperature, $\beta = c^2/v_{te}^2 \gg 1$, 
$\Delta = \omega^2 / \omega_{pe}^2$, $v_{te}^2 = T_e/m_e$, and 
$\omega_{pe}^2 = 4 \pi n_o q^2 / m_e$. Note that in our calculations, the character 
of the electromagnetic wave is determined by whether $\Delta$ is larger or smaller 
than unity. In the former case one has propagation in an under dense plasma and in 
the latter case one has propagation in an over dense plasma. Since it can be seen in 
Eq. (\ref{psinor}) that the coefficient of the linear $\psi$-term is given by 
$1-\Delta$, wave localization in over dense situations occurs when the natural 
frequency of the associated oscillator can be interpreted as an imaginary number. Our 
factor $\Delta$ is defined in terms of the effective frequency $\omega$ which 
incorporates all possible frequency shifts. Therefore this factor $\Delta$ replaces in 
a simplified way the slightly more complicated total frequency shift introduced in 
Ref. \cite{rao83}. This is why we adopt the present formalism. 
 
As for the ion acoustic field excited by ponderomotive effects associated with 
the electromagnetic wave, we simply write down the governing equation obtained from
the analysis of the low frequency dynamics involving the ion acoustic electric field 
potential $\Phi$, the massless warm electronic fluid, and the massive cold ionic fluid. 
One has \cite{rao83}
\begin{equation}
{d^2 \Phi \over d \xi^2} = - {M \over \sqrt{M^2 - 2 \, \Phi}} + 
e^{\Phi + \beta - \beta \, \sqrt{1+\psi^2}},
\label{finor}
\end{equation} 
where $\Phi$ has been normalized by $T_e / q$ ($q$ is the ion charge) and 
where we introduce the Mach number $M=V/C_s$ with the ion acoustic velocity $C_s$
written in the form $C_s = (T_e/m_i)^{1/2}$. The ions are considered nonrelativistic 
due to their large mass. We point out here that the total ion and electron 
densities are respectively measured by the absolute values of the first and second 
term on the right hand side of Eq. (\ref{finor}); in particular, the fluctuating 
electron density obtained under the assumption of massless electrons, 
\begin{equation}
\delta n_e = e^{\Phi + \beta - \beta \sqrt{1+\psi^2}} -1,
\label{elden}
\end{equation}
is the expression to be used in Eq. (\ref{psinor}).  

Equations (\ref{psinor}), (\ref{finor}) along with (\ref{elden}) govern the 
nonlinearly coupled dynamics of the dynamical variables $\psi(\xi)$ and $\Phi(\xi)$. 
We assume propagation at a constant velocity $M$ and take into consideration full 
nonlinear effects both in the ion acoustic and transverse relativistic dynamics of 
electrons. It is worth mentioning that alternative approaches do not restrict the 
spacetime dependence of the solutions but use weak nonlinear expansions instead 
\cite{moon90}-\cite{gio95}. 

Now, as shown in the paper by Rao {\it et al.} \cite{rao83}, both equations 
(\ref{psinor}) and (\ref{finor}) can be obtained from a generalized two degrees 
of freedom Hamiltonian $H$:
\begin{equation}
H = {P_{\psi}^2 \over 2 \beta^2} - {P_{\Phi}^2 \over 2} 
- {\Delta \beta \over 2} \psi^2 + M \sqrt{M^2 - 2 \Phi} + 
e^{\Phi + \beta - \beta\,\sqrt{1+\psi^2}}, 
\label{hamilton}
\end{equation}
where a misprint in \cite{rao83} has been corrected.
In the Hamiltonian, $P_\psi$ and $P_\Phi$ are the canonical momenta conjugate to 
the corresponding subscript coordinates. Hence, our system contains four dimensions: 
$\psi$, $P_\psi$, $\Phi$ and $P_\Phi$. 

Since the Hamiltonian (\ref{hamilton}) does not depend explicitly on the ``time'' 
coordinate $\xi$, it is a constant of motion. We will be interested in finding 
solitary waves and all those solutions towards which initial solitary waves can 
evolve in time if they are unstable. Therefore we shall work on the particular energy 
surface which allows for the presence of the configuration 
$\Phi = \psi = P_\Phi = P_\psi=0$, 
since this configuration is in fact the appropriate asymptotic solution for 
solitary pulses. 
We recall that from the canonical equations generated by $H$, Eq. (\ref{hamilton}), it follows 
$d\Phi/d\xi = - P_\Phi$ and $d\psi/d\xi = P_\psi/\beta^2$, and that the 
boundary conditions for a solitary pulse are, in the more traditional form, 
$\Phi,\>\>\psi,\>\>d\Phi/d\xi,\>\>d\psi/d\xi\rightarrow0$ as 
$|\xi|\rightarrow\infty$. The above considerations enable one to determine the 
constant numerical value of the Hamiltonian; it reads $H = 1 + M^2$. 

The Hamiltonian supports pure ion acoustic waves; it is easy to see that 
regardless the values of $P_\Phi$ and $\Phi$, if $\psi(\xi = 0) = P_\psi(\xi = 0)= 0$ 
then $\psi(\xi) = P_\psi(\xi)= 0$ at any $\xi$. This is not true for electromagnetic 
waves. Even if one starts with a laser pulse in the absence of any space charge 
fields ($\Phi = P_\Phi =0$), Eq. (\ref{finor}) indicates that the dynamics of the 
electric field potential is subsequently driven by a $\psi^2$ factor. 

We shall restrict the present analysis to subsonic cases, where $M < 1$. This causes 
the coefficient of the linear $\Phi$-term in Eq.(\ref{finor}) to assume only 
negative values. As for the laser field, we focus attention on over dense plasmas with 
$\Delta < 1$. Since the linear coefficient of Eq. (\ref{psinor}) is given by a factor 
$(1-\Delta)/\beta$, as already mentioned, the system is linearly unstable against the 
generation and propagation of electromagnetic modes. Nonlinear saturation of the 
unstable process may be responsible for the creation of solitary pulses.

\subsection{Adiabatic Approximation}

It has been seen that laser localization occurs in over dense plasmas where 
$\Delta < 1$. Since in general $\Delta$ is only slightly different from the unity 
and since the small factor $1-\Delta$ is yet to be divided by the large factor $\beta$ 
to obtain the coefficient of the linear term of Eq. (\ref{psinor}), the conclusion is 
that in general the following relation may hold:
\begin{equation}
\left({1-\Delta \over \beta}\right) \ll \left|1 - {1 \over M^2}\right|.
\label{ineq}
\end{equation}
But if such a relation does hold, it is likely that the dynamics on the $\psi,P_\psi$ 
phase plane tends to be much slower than the corresponding dynamics developing on the 
$\Phi,P_\Phi$ phase plane. In this limiting case one could be tempted to use the 
results of the center manifold and adiabatic theorems \cite{guc90} which say that 
the integration on the latter phase plane could be done simply by taking 
Eq. (\ref{finor}) with $\psi^2$ considered as a constant factor. In addition, as 
$\psi$ slowly evolves as a function of $\xi$, the $\Phi$-dynamics evolves in such a 
way as to conserve the action integral $(1/2\pi) \oint P_\Phi \, d\Phi$. In 
Fig. (\ref{fig_fips.ps}) we take fixed values of $\psi$ to plot contour levels of the 
driven ion acoustic Hamiltonian which is here defined as:
\begin{equation}
H_\Phi = - {P_\Phi^2 \over 2} + M \sqrt{M^2 - 2\,\Phi} + 
e^{\Phi + \beta - \beta \sqrt{1+\psi^2}}.
\label{hamio}
\end{equation}
In the adiabatic regime this is the Hamiltonian controlling the fast motion on 
the $\Phi,P_\Phi$ plane. For vanishingly small values of $\psi^2$ one can see in 
Fig. (\ref{fig_fips.ps}a) an elliptic fixed point at the origin and a hyperbolic point 
located at $\Phi < 0$; trajectories move counter-clockwise and the typical  
parameters $\beta = 100$ $M = 0.9$ are considered. Now as $\psi^2$ grows the 
elliptic point moves toward the hyperbolic point, as seen in Fig. (\ref{fig_fips.ps}b). 
For large enough values of $\psi$ the elliptic and hyperbolic fixed points 
coalesce in an inverse saddle-node bifurcation \cite{cor95}; we shall refer to this 
coalescence as collapse. For completeness we display in Fig. (\ref{fig_fips.ps}c) 
the $\psi,P_\psi$ phase space obtained from the full Hamiltonian (\ref{hamilton}) 
when we set $\Phi \rightarrow 0$ and $P_\Phi \rightarrow 0$. Note that when 
$\Delta<1$, as in Fig. (\ref{fig_fips.ps}c) where we consider $\Delta = 0.98$, the fixed point at the origin is hyperbolic; for $\Delta > 1$ the fixed point would be elliptic. 

A delicate point about the adiabatic approach is that the adiabatic theory may be 
expected to work relatively well only if the maximum value of $\psi^2$ throughout the 
entire dynamics is much smaller than the critical value for which the collapse does 
occur. Indeed, if this is the case the rotational frequency around the elliptic point 
on the $\Phi,P_\Phi$ plane can be expected to be larger enough than the characteristic 
time scale of the $\psi$-dynamics. But on the other hand if the maximum value of 
$\psi^2$ becomes too close to the critical value, the rotational frequency tends to 
diminish and attain values comparable to the $\psi$-time scales. The rotational 
frequency actually vanishes at the collapse. From this point of view, Eq. (\ref{ineq}) 
may not be sufficient to guarantee adiabaticity, since it is derived with basis on 
linearization procedures, where the fields are assumed to be much smaller than the 
maximum values they can actually attain as time elapses. With that in mind we now 
proceed to derive a validity condition for adiabaticity that takes into account the 
nonlinear effects associated with the finiteness of $\psi^2$. We shall see that the 
adiabatic range is in fact much smaller than the one suggested by Eq. (\ref{ineq}).  

Let us then examine the adiabatic trajectory of the elliptic point on the 
$\Phi, P_\Phi$ plane. Our interest lies on the fact that the existence condition for 
this point provides a reasonably good estimate on the range of validity of the 
adiabatic regime; we emphasize that adiabaticity is expected to break down when the 
elliptic and hyperbolic points cease to exist. One can use the results from the center manifold 
theorem to estimate the location of slowly moving fixed points, 
\begin{equation}
{\partial H_\Phi \over \partial \Phi}\>|_{fixed} = 
{\partial H_\Phi \over \partial P_\Phi}\>|_{fixed} = 0.
\label{fixedpoint}
\end{equation}
For a given value of $\psi$, one thus have 
\begin{equation}
e^{\Phi_{fixed} + \beta - \beta\,\sqrt{1+\psi^2}} - 
{M \over \sqrt{M^2 - 2\,\Phi_{fixed}}} = 0, 
\label{psifine} 
\end{equation}
from which a series expansion yields a relation correct up to quadratic terms 
\begin{equation}
\psi^2 = {2 \over \beta} \left(1-{1 \over M^2}\right) \, \Phi_{fixed} + 
{1-2 \, \beta - 2 \, M^2 + M^4 \over \beta^2 \, M^4} \Phi_{fixed}^2 \>;
\label{psifi}
\end{equation}
as for $P_\Phi$ one has $P_{\Phi,fixed} = 0$. We point out that in the extreme 
adiabatic limit where frequency shifts are vanishingly small, $\Delta \rightarrow 1$, 
the first term on the right hand side of Eq. (\ref{psifi}) coincides with expression 
(35) of Ref. \cite{rao83}. 

Given $\psi$, the quadratic relation above furnishes two roots in the variable $\Phi$ 
if the appropriate discriminant is positive. One of the roots represents the elliptic 
point, we shall call it $\Phi_{ell}(\psi^2)$, and the other represents the hyperbolic 
point, $\Phi_{hyp}(\psi^2)$, both seen in Figs. (\ref{fig_fips.ps}a) and 
(\ref{fig_fips.ps}b). What must be done now is to substitute the adiabatic relation 
$\Phi_{ell} = \Phi_{ell}(\psi^2)$ into Eq. (\ref{psinor}) to determine and examine the 
slow dynamics on the $\psi,P_\psi$ phase plane. Assuming for a moment $|\Phi|$ and 
$\psi^2$ small, which shall be seen to be true if $\Delta$ is sufficiently close to the 
unity, we drop the quadratic $\Phi$-term in Eq. (\ref{psifi}) and obtain 
\begin{equation}
\Phi_{ell} \approx {1 \over 2}\>{M^2\,\beta \over M^2 -1}\>\psi^2, 
\label{fiapro}
\end{equation}
which shows that $\Phi < 0$ if $M<1$. Therefore in the present approximation the 
$\psi$-dynamics is commanded by the following effective potential written, apart from 
a global multiplicative constant, as:
\begin{equation}
V_{eff}(\psi) = {1 \over 2} \, (\Delta-1 ) \, \psi^2 + 
{1 \over 8} \, \left(1 + \beta - {M^2\,\beta \over M^2 -1}\right) \, \psi^4.
\label{poefe}
\end{equation}
It is thus seen that near the ion acoustic resonance where $M \approx 1$, the 
electric field potential response satisfies $|\Phi| \gg \psi^2$ and essentially 
determines the adiabatically saturated value of the laser field:
\begin{equation}
\psi_{max}^2 \approx 4 \, {(\Delta - 1) \, (M^2 -1) \over \beta}.
\label{psimax}
\end{equation}
The approximation we have used in Eqs. (\ref{poefe}) and (\ref{psimax}) assumes that 
the quadratic $\Phi$-term in relation (\ref{psifi}) is much smaller than the others. This is 
true only if one is sufficiently away from that situation where the discriminant 
vanishes causing the collapsing of elliptic and hyperbolic points. Near the collapse, 
in particular, the adiabatic approximation is expected to break down. Given $M$ and 
$\beta$, use of Eq. (\ref{psimax}) and the self consistency requirement of a 
nonnegative discriminant for relation (\ref{psifi}) finally yields a complicated 
relation that can be used as an estimate for the critical value of $\Delta$ where 
the approximation (\ref{fiapro}) is no longer valid
\begin{equation}
{{{\Delta}} < \Delta_{cr} \equiv  
     {{{-\left( -4 + 7\,{\beta} + 8\,{M^2} + 
            {\beta}\,{M^2} - 4\,{M^4} \right) }\over 
        {4\,\left( 1 - 2\,{\beta} - 2\,{M^2} + {M^4}
             \right) }} \approx {7+M^2 \over 8}}}.
\label{delcri}
\end{equation}
As $\Delta$ starts to get too close to $\Delta_{cr}$ the adiabatic approximations 
are expected to get poorer and poorer. Condition (\ref{delcri}) is a rough estimate 
which could be refined with more detailed algebraic work. However we shall take 
it as sufficiently accurate and complete for our purposes. In any case the ultimate 
answer is yet to be given by numerical work as we will do next.

\section{Transition from Adiabatic to Chaotic Regimes}  

\subsection{Testing Adiabaticity} 

In all the following numerical applications we use $\beta = 100$ and $M=0.9$ 
which yields $\Delta_{cr} \approx 0.976$. We promptly 
conclude that the validity range for the adiabatic regime is quite narrow, as a 
matter of fact, much narrower than the range predicted by Eq. (\ref{ineq}). Indeed, 
for the chosen values of $\beta$ and $M$, and considering over dense plasmas, 
Eq. (\ref{ineq}) basically imposes no essential restriction on the value of $\Delta$, 
a failure of the linear theory as mentioned before. In the following we shall see that 
the estimate based on $\Delta_{cr}$, Eq. (\ref{delcri}), is much more accurate than 
the one based on Eq. (\ref{ineq}), and that the destruction of the adiabatic regime is 
in fact associated with a transition to chaos.

Before examining the validity ranges and the transition, let us first perform some 
initial simulations of Eqs. (\ref{psinor}) and (\ref{finor}) to make sure that the 
adiabatic regime is in fact present if condition (\ref{delcri}) is safely observed. To 
do so we start a single initial condition with $\Delta = 0.99$, and with 
$\psi = P_\Phi = 0$, $\Phi=0.0005$ and $P_\psi = 0.0024$ such that $H = 1+M^2$. We 
plot the time series for $\psi(\xi)$ and $\Phi(\xi)$ in Fig. (\ref{fig_timese.ps}). In 
the figure we see that $\Phi$ undergoes a fast oscillatory motion while $\psi$ evolves 
in a much slower time scale. The adiabatic features can be also visualized on the 
$\psi,P_\psi$ phase plane as in Fig. (\ref{fig_psips.ps}). In the figure we compare 
three solitary trajectories: the exact trajectory, the adiabatic trajectory calculated 
from Eq. (\ref{psinor}) under the assumption (\ref{fiapro}), and the 
trajectory calculated from Eq. (\ref{psinor}) under the assumption 
$\Phi \rightarrow 0$. The adiabatic trajectory yields a fairly good 
approximation to the actual trajectory. Here we use $\Delta = 0.98$. This chosen value 
of $\Delta$ is slightly smaller than in the previous figure, because it allows a 
clearer view of the differences between adiabatic and exact trajectories. For values of 
$\Delta$ closer to the unity the approximation gets better and better until such a 
point where no distinction can be appreciated. 

\subsection{The Transition}

Now the question refers to what happens as the parameters are varied beyond the 
validity range for the adiabatic regime. To simplify the discussion we shall focus 
attention on the behavior of the system as $\Delta$ decreases. As our system is 
Hamiltonian (see Eq. (\ref{hamilton})) with two degrees of freedom, we make use of the 
Poincar\'e map methodology and plot the pair of phase variables $\Phi$ and $P_\Phi$ 
each time $P_\psi = 0$ with $dP_\psi/d\xi > 0$. Several initial conditions are 
launched with the numerical values for $\Phi(\xi = 0)$ distributed within a small 
range typically satisfying $-0.001 < \Phi(\xi = 0) < 0.001$. Similarly to the initial 
simulation presented before, for all initial conditions we always take 
$\psi = P_\Phi=0$ and calculate the corresponding initial $P_\psi$ from the constant 
numerical value of the Hamiltonian, $H = 1 + M^2$. This kind of launching conditions 
initially places the system in the vicinity of the solitary solution which is the 
solution containing the point $\Phi = P_\Phi = \psi = P_\psi = 0$. In integrable cases 
the ensuing nearly solitary trajectories progress in fact as trains of solitons, but 
even in the nonintegrable cases where solitons are not seen, the trajectories still 
cross the $P_\psi = 0$ plane several times, an essential condition for the 
construction of the maps. 

We start by displaying in Fig. (\ref{fig_plot.ps}a) the map obtained when 
$\Delta = 0.98$. For such a value of $\Delta$ the adiabatic regime is expected 
to prevail. In agreement with that, what is seen in the plot is a set of regular 
orbital concentric curves. Note that the elliptic point appears to be located at 
a negative value of $\Phi$ simply because this is the value of the electric field 
potential when the recording conditions $P_\psi = 0$, $dP_\psi/d\xi > 0$ are 
satisfied. Now if one starts to decrease $\Delta$ the transition to 
chaos is expected to occur. Let us move on to Fig. (\ref{fig_plot.ps}b) where 
$\Delta = 0.975$. As anticipated from the analytical estimates, a considerable amount 
of chaotic activity can already be identified. The central region of the map is 
completely surrounded by a blend of stochastic orbits and resonant islands. In 
particular, it appears that the soliton solution which corresponds to the central 
fixed point no longer exists. In Fig. (\ref{fig_plot.ps}c) we enlarge part of 
Fig. (\ref{fig_plot.ps}b) to show details of the resonant islands. These small 
remaining regions of regularity of the phase space are then totally suppressed when 
one reduces $\Delta$ further below. In Fig. (\ref{fig_plot.ps}d), for instance, we 
consider the case $\Delta = 0.97$ to show a deep chaotic regime. In conclusion, a 
complete destruction of the regular trapping region does indeed occur as $\Delta$ 
decreases. But if there is no trapping region, how would the trajectories behave? We 
will address this issue next.   

\subsection{Asymptotic States Beyond the Transition} 

We can actually see some persistency in chaos only for those intermediary situations 
where $\Delta$ is not too close to the unity, but not too small. Indeed, if $\Delta$ 
becomes sufficiently small, say $\Delta < \Delta_{cr}$, no trapping region is 
effectively formed on the $\Phi,P_\Phi$ plane. Even initial conditions originally 
launched within the trapping region predicted by the adiabatic theory when $\psi = 0$, 
do not remain there. As a matter of fact, the trajectories are eventually ejected into 
unbounded regions of the phase space when $\Delta$ crosses the validity limits of 
adiabaticity. For those cases, chaos would be at most a transient which would take 
place during initial instants, before ejection. The question to be asked now should be 
on the configuration of these unbounded orbits. What we have observed is that once the 
trajectory escapes from the trapping region in the ion acoustic phase space, it starts 
to follow the open flow lines of Fig. (\ref{fig_fips.ps}). This is confirmed in Fig. 
(\ref{fig_escape.ps}a) where we show a continuous plot displaying a freed trajectory 
which was started with $\Delta=0.97$ and $\Phi(\xi=0)=0.0001$ (figures 
(\ref{fig_escape.ps}a) and (\ref{fig_escape.ps}b) are not Poincar\'e maps; the 
trajectory points are periodically recorded with a small but constant time step). The 
fact that the subsequent trajectory evolves along the flow lines implies that $\Phi$ 
gets more and more negative. Now, if one considers Eq. (\ref{hamilton}) one readily 
sees that regardless the value of $\psi$, 
\begin{equation}
\lim_{\Phi \rightarrow - \infty} H = {P_\psi^2 \over 2 \, \beta^2} + 
{\Delta \, \beta \over 2} \psi^2 - {P_\Phi^2 \over 2} + M\,\sqrt{M^2-2\,\Phi}.
\label{hamfree}
\end{equation}
In other words, laser and ion acoustic fields become decoupled in this limit. Since 
$\Delta\beta>0$ then the corresponding dynamics of the laser field must 
necessarily become that of an undriven harmonic oscillator. This is what is shown in 
Fig. (\ref{fig_escape.ps}b) where we project the same dynamics of 
Fig. (\ref{fig_escape.ps}a) now on the $\psi,P_\psi$ phase plane. After a certain 
amount of time following a figure eight shape like the ones seen in 
Figs. (\ref{fig_fips.ps}c) and (\ref{fig_psips.ps}), there is a dynamical transition 
to the circular shape so characteristic of the harmonic oscillator. The instant of the 
transition coincides, as it should, with the moment of ejection seen in 
Fig. (\ref{fig_escape.ps}a), and occurs approximately after $30$ 
cycles of the laser wave in its initial figure eight phase trajectory. Another 
interesting point connected to this asymptotic state is that as 
$\Phi\rightarrow - \infty$, the particle density becomes very small 
(see Eq. (\ref{elden})). Noticing that $\Phi < 0$, the ponderomotive field created by 
the laser induces an initial potential well in which interior the ion fluid undergoes 
acceleration becoming less dense. If the amplitude of the laser is too large the 
process is unstable and never arrests. When the density is low the laser becomes an 
almost standing wave with a small propagation velocity of the crests ($V \ll c$).  

\subsection{The Role of Relativistic Effects}

Our original equation, Eq. (\ref{psinor}), includes full relativistic electronic 
nonlinearities. As a final topic it is perhaps interesting to discuss the role of 
these relativistic nonlinear effects as compared to ponderomotive nonlinear effects. 

\begin{figure}
\caption{Influence of relativistic nonlinearities on the dynamics. In the lower panel 
of each figure relativistic effect is artificially suppressed. In (a) we consider an 
adiabatic regular regime with $\Delta=0.98$ and in (b) we take $\Delta = 0.97$ to 
produce an escaping trajectory after some transient chaos. It is seen that in the 
chaotic case one should not discard relativistic effects.}
\label{fig_relati.ps}
\end{figure}
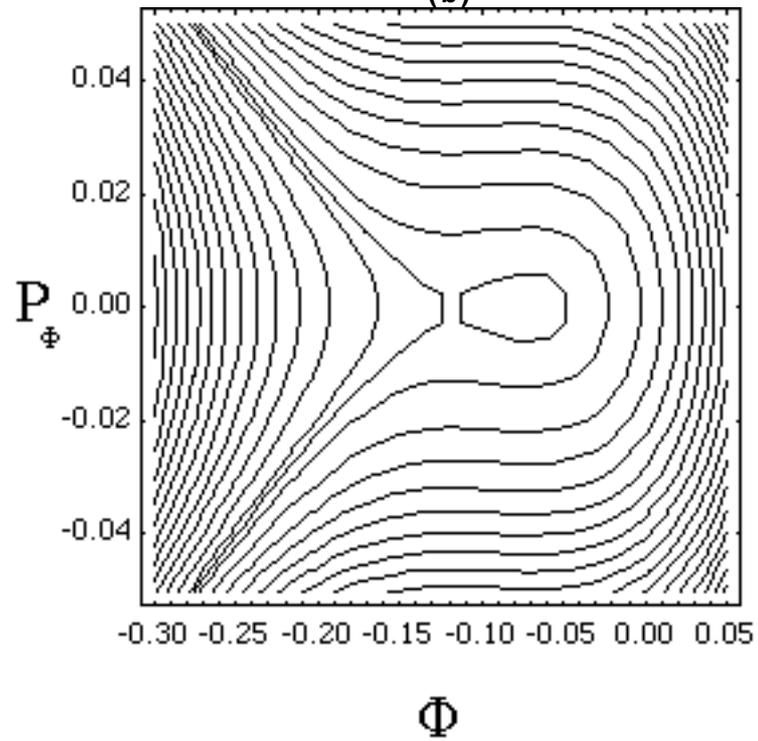
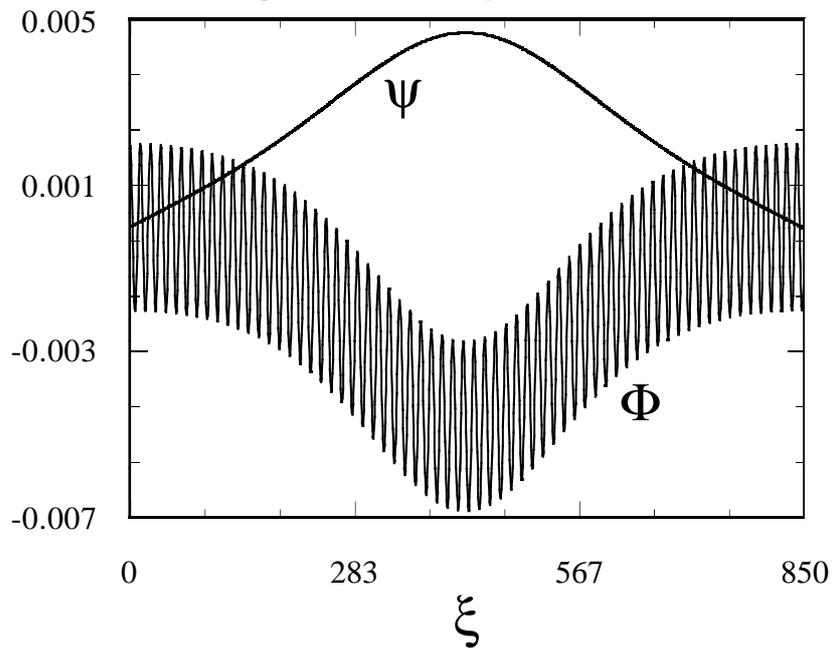
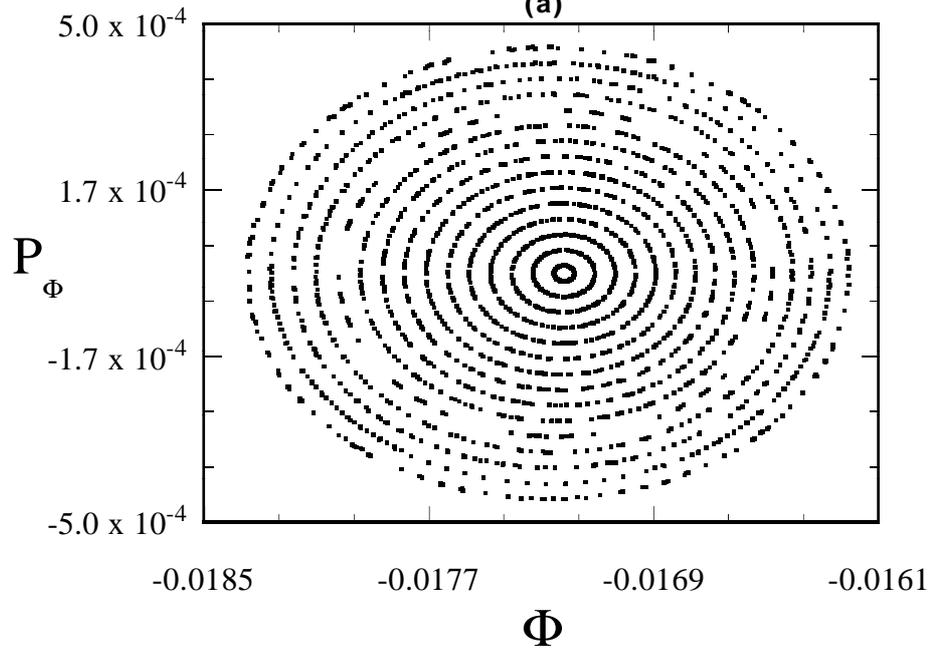
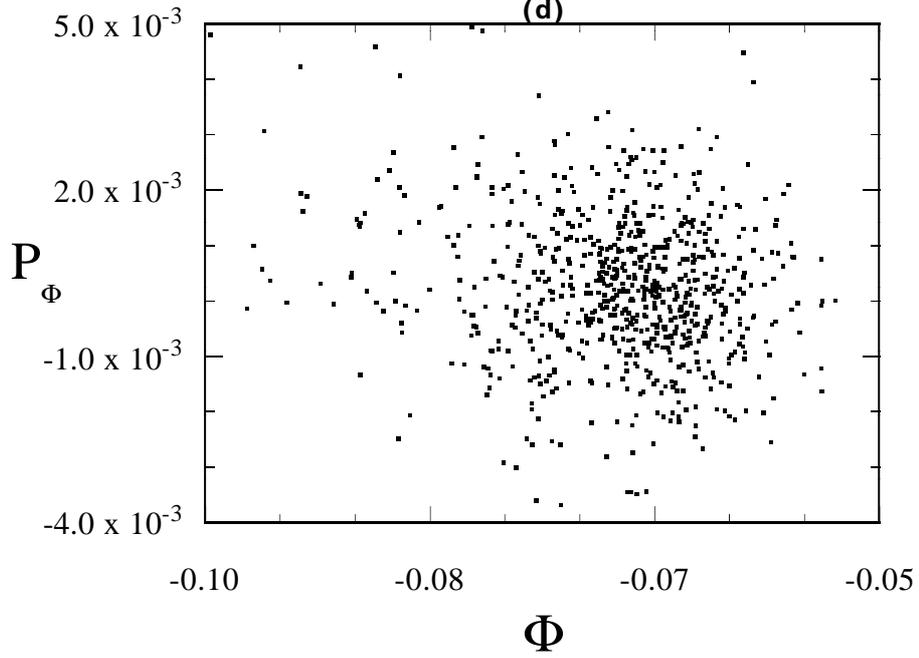
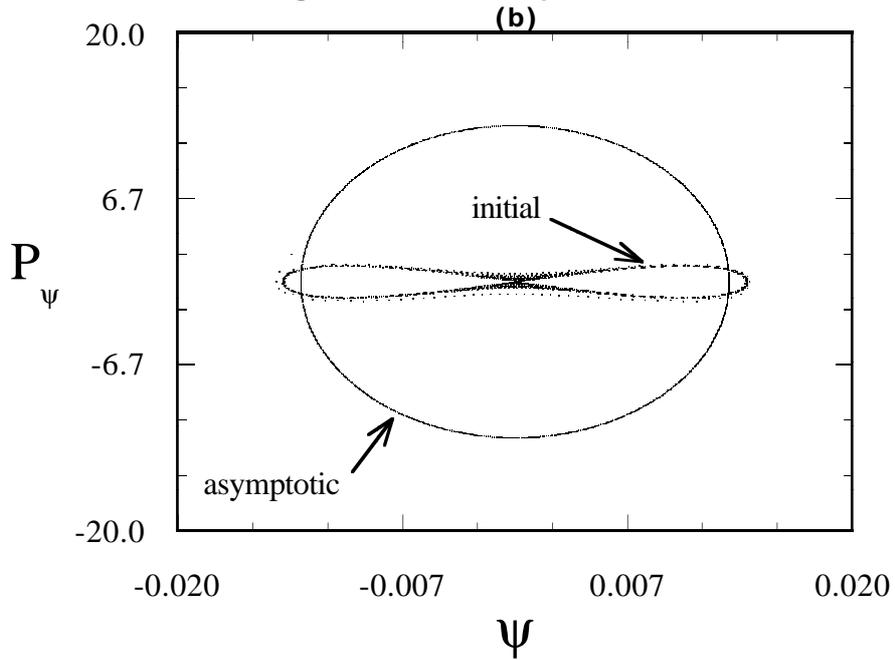
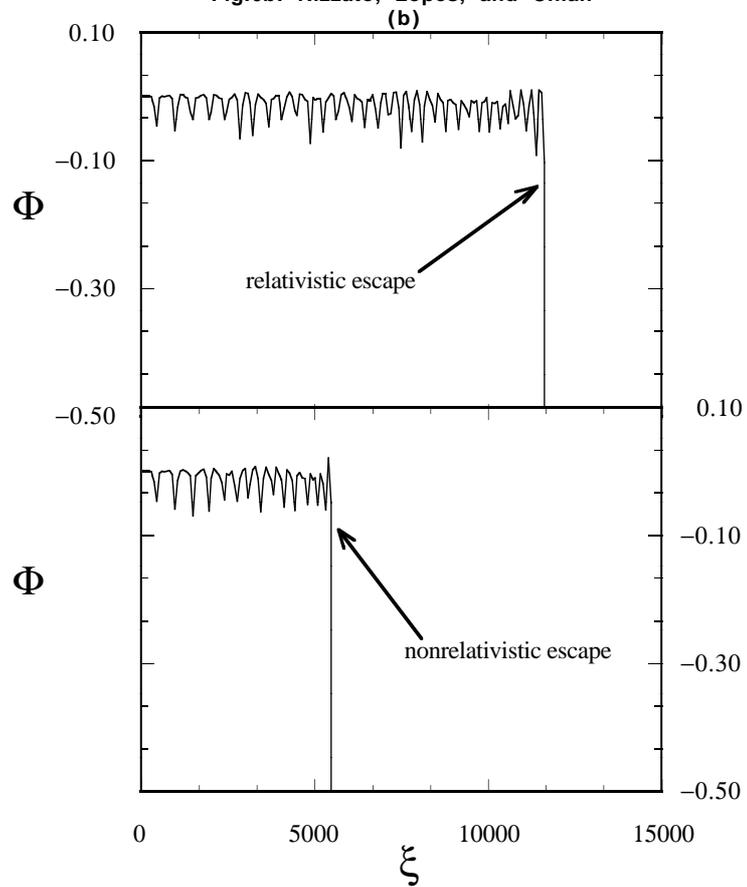

What we find here is that while in the adiabatic regime, saturation is essentially 
governed by ponderomotive nonlinearities. This is the basic conclusion associated 
with Eq. (\ref{psimax}).  Now we would like to know whether or not relativistic 
effects grow in importance in chaotic regimes. To this end we perform 
two pair of simulations that are displayed in Fig. (\ref{fig_relati.ps}). In the upper 
panel of each pair we depict time series for $\Phi(\xi)$ considering exact fully 
relativistic nonlinear dynamics. In the lower panel we consider the time series with 
relativistic mass correction suppressed. It is seen that in the adiabatic regime of 
Fig. (\ref{fig_relati.ps}a) where we consider $\Delta = 0.98$, relativistic effects 
are not prominent, as both figures are almost identical. On the other hand, in 
Fig. (\ref{fig_relati.ps}b) where we consider $\Delta=0.97$ relativistic effects are 
of relevance. Indeed, in the time series of Fig. (\ref{fig_relati.ps}b) we see that 
for the particular initial conditions we use, the nonrelativistic trajectory undergoes 
much earlier ejection from the initial trapping region. For other initial conditions, 
the escape order may be reversed and the nonrelativistic trajectory may be ejected 
after the relativistic one. What is really remarkable here, however, is that even in 
our case of small field amplitudes where relativistic effects are small, relativistic 
and nonrelativistic trajectories may largely differ in chaotic regimes, the reason for 
this being the extreme sensibility of chaotic systems to parameter variations. 

\section{Final Conclusions}

We have performed a nonlinear analysis on the interaction of high power laser 
waves with ion acoustic waves in a plasma. We assume stationary propagation in an 
over dense plasma and consequently show that three generic configurations take 
place. If the effective laser frequency is only slightly lower than the average electron 
plasma frequency, the ensuing dynamics is adiabatic. The faster varying ion acoustic 
field is adiabatically enslaved to the slowly varying envelope laser field, and the 
resulting electromagnetic envelope solitons are likely to exist and remain stable. As 
one starts to decrease the effective laser frequency, adiabaticity becomes progressively 
poorer. A blend of confined chaotic regions and nonlinear resonance islands are seen 
on the $\Phi,P_\Phi$ phase space. Then, for yet smaller values of the laser frequency 
adiabaticity is completely destroyed. Initial conditions are rapidly ejected from the 
trapping region on the ion acoustic phase space and proceed to move along unbounded 
curves towards $\Phi \rightarrow - \infty$. In this limit the system becomes decoupled, 
the laser field starts to behave like a vacuum field and the plasma becomes 
progressively rarefied. 

Relativistic effects are moderate in adiabatic regimes but of considerable relevance in 
chaotic regimes. We saw that trajectories with relativistic corrections artificially 
removed can largely differ from the exact ones, the reason being the sensibility of 
chaotic systems to parameter variations.

One general conclusion obtained here is that the parameter range for the existence of 
solitary wave is extremely narrow. This numerically confirmed fact can be predicted 
only by nonlinear estimates like that provided by Eq. (\ref{delcri}), which takes 
into account the effect of finite values of $\psi^2$ on the existence of 
fixed points. Linear estimates like that of Eq. (\ref{ineq}) are much less 
accurate. Due to this narrow existence range, we have not observed, for instance, 
double hump solitons like those obtained in Ref. \cite{rao83}. The point is that 
according to the calculations done in this reference, and making the appropriate 
connections and translations between the various formulas, double hump solitons exist 
only when $\Delta$ reaches small values, $\Delta \sim 0.7$. However, for such a small 
values of this parameter, our system has already lost stability due to the transition 
to chaos.   

\centerline{\bf ACKNOWLEDGMENTS}

This work was partially supported by 
Financiadora de Estudos e Projetos (FINEP) and Conselho Nacional de 
Desenvolvimento Cient\'{\i}fico e Tecnol\'ogico (CNPq), Brazil. Numerical 
computing was performed on the CRAY Y-MP2E at the Universidade Federal do 
Rio Grande do Sul Supercomputing Center.  

\newpage

%

%
%
\begin{figure}
\caption{Phase space on the $\Phi,P_\Phi$ plane for fixed values of $\psi$, with 
$M=0.9$ and $\beta=100$; $\psi=0$ in (a), and $\psi = 0.001$ in (b). In (c) we 
display the $\psi,P_\psi$ phase space using $\Delta = 0.98$.}
\label{fig_fips.ps}
\end{figure}
%
\begin{figure}
\caption{Time series in the adiabatic regime; $M=0.9,\,\beta=100,\,
\Delta=0.99$}
\label{fig_timese.ps}
\end{figure}
%
\begin{figure}
\caption{Accuracy of the adiabatic approximation tested on the $\psi,P_\psi$ plane; 
$M=0.9$, $\beta=100$ and $\Delta = 0.98$.}
\label{fig_psips.ps}
\end{figure}
%
\begin{figure}
\caption{Transition to chaos for $M=0.9$ and $\beta=100$; $\Delta=0.98$ in (a), 
$\Delta=0.975$ in (b), and $\Delta=0.97$ in (d). In (c) we show details of the 
resonant islands seen in (b).}
\label{fig_plot.ps}
\end{figure}
%
\begin{figure}
\caption{Asymptotic states on the phase planes. The escaping trajectory seen 
in (a) and the circular trajectory seen in (b) do appear on the respective phase 
planes after approximately $10$ figure eight cycles of the $\psi,P_\psi$ variables; 
$M=0.9,\,\beta=100,\,\Delta=0.97$}
\label{fig_escape.ps}
\end{figure}  
%
\end{document}